\documentclass{article}

\usepackage{graphicx}
\usepackage{geometry}
\usepackage{amsmath}
\usepackage{amssymb}
\usepackage{tikz}
\usetikzlibrary{decorations.shapes}
\usepackage{authblk}
\usepackage{setspace}
\usepackage{stmaryrd}
\usepackage{algorithm}
\usepackage{algpseudocode}
\usepackage{listings,xcolor}
\usepackage{inconsolata}

\begin{document}

\title{Generation and analysis of lamplighter programs}

\author{Carlos Martin}

\affil{Department of Computer Science,
  Columbia University\\
  2920 Broadway,
  New York, NY 10027,
  United States\\
  \texttt{carlos.martin@columbia.edu}
}

\maketitle

\begin{abstract}
    We consider a programming language based on the lamplighter group that uses only composition and iteration as control structures. We derive generating functions and counting formulas for this language and special subsets of it, establishing lower and upper bounds on the growth rate of semantically distinct programs. Finally, we show how to sample random programs and analyze the distribution of runtimes induced by such sampling.
\end{abstract}


\section{Introduction} \label{introduction}

The structured program theorem states that every computable function can be implemented in a programming language using only composition and iteration as control structures \cite{Bohm1964}\cite{Bohm1966}. In this paper, we study a programming language based on P\(''\) that manipulates an infinite sequence of bits with the following primitive instructions:
\begin{center}
\begin{tabular}{l l}
    \(t\) & toggle current bit \\
    \(r\) & move right \\
    \(l\) & move left \\
    \([E]\) & repeat \(E\) while current bit is 1
\end{tabular}
\end{center}

We occasionally use the notation \texttt{+}, \texttt{>}, and \texttt{<} for the first three instructions, in order.

More complex instructions can be created by combining these. For example, \([t]\) sets the current bit to 0, \([t]t\) sets it to 1, and \([]t[]\) loops forever unconditionally. \(t[tEt]t\) repeats \(E\) while the current bit is 0, inverting the loop condition. The state of the current bit is preserved because either \(t\) preceding a check for the loop condition is immediately cancelled by a subsequent \(t\).

A more complex instruction is \([tr[r]r[r]trt[l]l[l]r]\), which doubles a string of 1s using two buffers as described below. It starts at the leftmost 1 and transforms \(1^k 0 0^{2k}\) into \(0^k 0 1^{2k}\). Note that the location of the doubled string lies to the right of the initial string of 1s.
\begin{center}
\begin{tabular}{l l}
    \(t\) & remove 1 tally mark from left buffer \\
    \(r[r]r[r]\) & move to right buffer \\
    \(trt\) & add 2 tally marks to right buffer \\
    \([l]l[l]r\) & move to left buffer
\end{tabular}
\end{center}

The outer loop stops when the left buffer has been exhausted. The execution of this instruction on the input \(1^2 0 0^4\) is shown below:
\begin{align*}
    \ldots \underline{\blacksquare} \blacksquare \square \square \square \square \square \ldots \\[-.5em]
    \ldots \underline{\square} \blacksquare \square \square \square \square \square \ldots \\[-.5em]
    \ldots \square \underline{\blacksquare} \square \square \square \square \square \ldots \\[-.5em]
    \ldots \square \blacksquare \underline{\square} \square \square \square \square \ldots \\[-.5em]
    \ldots \square \blacksquare \square \underline{\square} \square \square \square \ldots \\[-.5em]
    \ldots \square \blacksquare \square \underline{\blacksquare} \square \square \square \ldots \\[-.5em]
    \ldots \square \blacksquare \square \blacksquare \underline{\square} \square \square \ldots \\[-.5em]
    \ldots \square \blacksquare \square \blacksquare \underline{\blacksquare} \square \square \ldots \\[-.5em]
    \ldots \square \blacksquare \square \underline{\blacksquare} \blacksquare \square \square \ldots \\[-.5em]
    \ldots \square \blacksquare \underline{\square} \blacksquare \blacksquare \square \square \ldots \\[-.5em]
    \ldots \square \underline{\blacksquare} \square \blacksquare \blacksquare \square \square \ldots \\[-.5em]
    \ldots \underline{\square} \blacksquare \square \blacksquare \blacksquare \square \square \ldots \\[-.5em]
    \ldots \square \underline{\blacksquare} \square \blacksquare \blacksquare \square \square \ldots \\[-.5em]
    \ldots \square \underline{\square} \square \blacksquare \blacksquare \square \square \ldots \\[-.5em]
    \ldots \square \square \underline{\square} \blacksquare \blacksquare \square \square \ldots \\[-.5em]
    \ldots \square \square \square \underline{\blacksquare} \blacksquare \square \square \ldots \\[-.5em]
    \ldots \square \square \square \blacksquare \underline{\blacksquare} \square \square \ldots \\[-.5em]
    \ldots \square \square \square \blacksquare \blacksquare \underline{\square} \square \ldots \\[-.5em]
    \ldots \square \square \square \blacksquare \blacksquare \underline{\blacksquare} \square \ldots \\[-.5em]
    \ldots \square \square \square \blacksquare \blacksquare \blacksquare \underline{\square} \ldots \\[-.5em]
    \ldots \square \square \square \blacksquare \blacksquare \blacksquare \underline{\blacksquare} \ldots \\[-.5em]
    \ldots \square \square \square \blacksquare \blacksquare \underline{\blacksquare} \blacksquare \ldots \\[-.5em]
    \ldots \square \square \square \blacksquare \underline{\blacksquare} \blacksquare \blacksquare \ldots \\[-.5em]
    \ldots \square \square \square \underline{\blacksquare} \blacksquare \blacksquare \blacksquare \ldots \\[-.5em]
    \ldots \square \square \underline{\square} \blacksquare \blacksquare \blacksquare \blacksquare \ldots \\[-.5em]
    \ldots \square \underline{\square} \square \blacksquare \blacksquare \blacksquare \blacksquare \ldots \\[-.5em]
    \ldots \square \square \underline{\square} \blacksquare \blacksquare \blacksquare \blacksquare \ldots
\end{align*}

In this paper, we study both the syntax and semantics of the programming language generated by the primitive instructions listed above. The primary goal is to enumerate programs that are behaviorally unique by finding classes of programs that are semantically equivalent. To this end, we study the relationship between this programming language and the lamplighter group, as well as various loop reachability conditions, invariants, and transformations.

To measure progress, we analyze the asymptotic behaviour of the generating functions associated with these increasingly-refined subsets of the original language. This establishes increasingly tight upper and lower bounds on the growth rate of semantically distinct programs.

\subsection{Generating grammar}

The programming language is generated by the grammar
\[ E = (t + r + l + [E])^* \]

where \(E\) is the start symbol. Therefore, its generating function is the solution to
\begin{align*}
    E(z) &= (z + z + z + z^2 E(z))^* \\
    &= \frac{1}{1 - (3z + z^2 E(z))}
\end{align*}

which is
\begin{align*}
    E(z) &= \frac{2}{1 - 3z + \sqrt{(1 - z)(1 - 5z)}} \\
    &= 1 + 3z + 10z^2 + 36z^3 + 137z^4 + 543z^5 + 2219z^6 + 9285z^7 + \ldots
\end{align*}

where the \(n\)th coefficient, denoted by \([z^n] E(z)\), is the number of expressions of length \(n\). Since \(E(z)\) has branch points at \(z = 1\) and \(z = \frac{1}{5}\), its radius of convergence is \(\frac{1}{5}\). Therefore \cite{Flajolet}
\[ \lim_{n \rightarrow \infty} ([z^n] E(z))^{1/n} = 5 \]

Now, suppose a function \(f\) takes the following form with \(\alpha, \beta \in \mathbb{R}\):
\[ f(z) = \frac{A(z) + B(z) \left( 1 - \frac{z}{r} \right)^{-\beta}}{z^\alpha} \]

where \(r\) is the radius of convergence of \(f\), the radius of convergence of \(A\) and \(B\) is greater than \(r\), and \(B(r) \neq 0\). Then an asymptotic expression for its coefficients is given by \cite{Flajolet}
\[ [z^n] f(z) \sim \frac{B(r)}{r^\alpha \Gamma(\beta)} n^{\beta-1} r^{-n} \]

In particular,
\[ \lim_{z \rightarrow r} \frac{f(z) \left(1 - \frac{z}{r}\right)}{\Gamma(\beta)}^\beta = \lim_{z \rightarrow r} \frac{A(z) (1 - \frac{z}{r})^\beta + B(z)}{z^\alpha \Gamma(\beta)} = \frac{B(r)}{r^\alpha \Gamma(\beta)} \]

yields the asymptotic coefficient if \(\beta > 0\). For \(E(z)\) we have \(r = \frac{1}{5}\), \(\alpha = 2\), \(\beta = -\frac{1}{2}\), and
\begin{align*}
    A(z) &= \frac{1 - 3z}{2} \\
    B(z) &= -\frac{(1 - z)^{1/2}}{2}
\end{align*}

Thus an asymptotic expression for \([z^n] E(z)\) is
\[ [z^n] E(z) \sim \frac{5^{n + 3/2}}{2 \sqrt{\pi n^3}} \]

Notice that the instructions \(t\), \(r\), and \(l\) generate a group under composition. We can reduce the number of equivalent instruction sequences by using one representation for each group element. This group is called the lamplighter group, and is described in the next section.
\section{The lamplighter group} \label{lamplighter}

The lamplighter group \(\mathcal{L}\) is the restricted wreath product
\[ \mathcal{L} = (\mathbb{Z}/2\mathbb{Z}) \wr \mathbb{Z} = (\mathbb{Z}/2\mathbb{Z})^{\oplus \mathbb{Z}} \rtimes \mathbb{Z} \]

where \(\oplus\) is the direct sum and \(\rtimes\) is the semidirect product. Its standard presentation is
\[ \langle t, r \mid t^2, [r^k t r^{-k}, r^m t r^{-m}] \text{ for } k, m \in \mathbb{Z} \rangle \]

where \(r^{-1} = l\). The group can be viewed as the action of a lamplighter on an infinite sequence of lamps. \(t\) toggles the lamp above the lamplighter, \(r\) moves the lamplighter right, and \(l\) moves the lamplighter left. Each group element is an ordered pair \((a, b)\) where \(a \in (\mathbb{Z}/2\mathbb{Z})^{\oplus \mathbb{Z}}\) and \(b \in \mathbb{Z}\). The group operation is
\[ (a_1, b_1) (a_2, b_2) = (a_1 + \sigma^{b_1} a_2, b_1 + b_2) \]

and the group inverse is
\[ (a, b)^{-1} = (-\sigma^{-b} a, -b) \]

where \(\sigma\) is the shift operator, defined by \((\sigma a)(i) = a(i+1)\). For example, the illustration
\[ \ldots \square \square \blacksquare \square \overset{\downarrow}{\square} \blacksquare \square \underline{\square} \square \ldots \]

represents the element \((\delta_{-2} + \delta_{1}, 3)\), where \(\delta_i \in (\mathbb{Z}/2\mathbb{Z})^{\oplus \mathbb{Z}}\) is defined by
\begin{align*}
    \delta_i(j) = \begin{cases}
        1 & i = j \\
        0 & \text{otherwise}
    \end{cases}
\end{align*}

In this illustration, the arrow points to the lamp at the origin, black squares indicate toggled lamps, and the underscore points to the final position of the lamplighter. The group identity of the lamplighter group is \((0, 0)\). A generating set is given by \(\{t, r, l\}\) where
\begin{align*}
    t &= (\delta_0, 0) \\
    r &= (0, 1) \\
    l &= (0, -1)
\end{align*}

\subsection{Word norm}

Let \(G\) be a group and \(S \subseteq G\) be a generating set of \(G\). A word \(w \in S^*\) is a string of generators whose product \(\llbracket \cdot \rrbracket : S^* \rightarrow G\) under the group operation is an element of \(G\):
\begin{align*}
    \llbracket (s_1, s_2, \ldots, s_n) \rrbracket &= s_1 s_2 \ldots s_n
\end{align*}

The empty word \(\varepsilon\) represents the empty product, which is the group identity. The word norm \(\lvert \cdot \rvert : G \rightarrow \mathbb{N}\) of a group element \(g \in G\) is the minimum length of any word that represents it:
\begin{align*}
    \lvert g \rvert = \min\, \{ \lvert w \rvert \mid w \in S^*, \llbracket w \rrbracket = g \}
\end{align*}

The norm of the group identity is zero since it can be represented as \(\varepsilon\). The word norm of a lamplighter group element \((a, b)\) under the generating set \(\{t, r, l\}\) is \cite{Cleary2005}
\[ \lvert (a, b) \rvert = \lVert a \rVert_1 + a_{\uparrow} - a_{\downarrow} + \begin{cases}
    \lvert a_{\downarrow} \rvert + \lvert a_{\uparrow} - b \rvert & b \geq 0 \\
    \lvert a_{\uparrow} \rvert + \lvert a_{\downarrow} - b \rvert & \text{otherwise}
\end{cases} \]

where \(\lVert \cdot \rVert_1\) is the 1-norm
\begin{align*}
    \lVert a \rVert = \sum_{k \in \mathbb{Z}} \lvert a(k) \rvert
\end{align*}

and
\begin{align*}
    a_{\downarrow} &= \begin{cases}
        0 & a = 0 \\
        (\min \circ \operatorname{supp})(a) & \text{otherwise}
    \end{cases} \\
    a_{\uparrow} &= \begin{cases}
        0 & a = 0 \\
        (\max \circ \operatorname{supp})(a) & \text{otherwise}
    \end{cases}
\end{align*}

This can be seen by dividing the movement of the lamplighter into three stages, as shown in the following diagram:
\begin{center}
\begin{tikzpicture}
    \fill [lightgray] (0,0) rectangle (.25,5.25);
    \fill [lightgray] (1,0) rectangle (1.25,5.25);
    \fill [lightgray] (2,0) rectangle (2.25,5.25);
    \fill [lightgray] (3,0) rectangle (3.25,5.25);
    \draw[gray,step=.25cm] grid (3.25,5.25);
    \begin{scope}[shift={(1.125,5.125)}]
        \draw [->,>=latex] (0, 0) -- (-1, -1) -- (2, -4) -- (1, -5);
        \node at (-1, .35) {\(a_\downarrow\)};
        \node at (0, .35) {\(0\)};
        \node at (1, .35) {\(b\)};
        \node at (2, .35) {\(a_\uparrow\)};
    \end{scope}
\end{tikzpicture}
\end{center}

A minimal-length word representing a lamplighter group element is a solution to the traveling salesman problem in one dimension, with the locations of the toggled lamps as vertices. If \(b \geq 0\), the lamplighter visits the leftmost toggled lamp, the rightmost toggled lamp, and \(b\) in that order. Similarly, if \(b < 0\), the lamplighter visits the rightmost toggled lamp, the leftmost toggled lamp, and \(b\) in that order. The set of all lamplighter group elements with \(b \geq 0\) is given by
\[ \underbrace{(l^i(g-1)r(gr)^{i-1})^{[i > 0]}}_\text{stage 1} \quad \underbrace{g(rg)^j}_\text{stage 2} \quad \underbrace{(r^k(g-1)l(gl)^{k-1})^{[k > 0]}}_\text{stage 3} \]

for all \(i, j, k \in \mathbb{N}\), where \(g = \varepsilon + t\) and \([\cdot]\), in this context, denotes the Iverson bracket. Similarly, the set of all elements with \(b < 0\) is given by the expression above with \(l\) and \(r\) switching places. This representation is a minimal-length normal form for each lamplighter group element. It maps each group element \(g \in G\) to a word \(w \in S^*\) that represents it.

The word distance \(d : G \times G \rightarrow \mathbb{N}\) between two elements is the smallest number of generators needed to change one into the other. For a symmetric generating set, it can be expressed as
\[ d(g, h) = \lvert g^{-1} h \rvert \]

\subsection{Growth function}

The growth function, here denoted by \(G(z)\), of a group \(G\) is a formal power series whose \(n\)th coefficient is the number of group elements with a word norm of \(n\). For example,
\[ (\mathbb{Z}/n\mathbb{Z})(z) = \sum_{0 \leq k < n} z^k = \frac{1 - z^n}{1 - z} \]

with respect to the singleton generating set, and
\[ \mathbb{Z}(z) = \sum_{k \in \mathbb{Z}} z^{\lvert k \rvert} = \frac{1 + z}{1 - z} \]

with respect to the standard generators \(\{1, -1\}\). The growth function of \(G \wr \mathbb{Z}\), where \(\wr\) is the restricted wreath product, can be expressed in terms of \(G(z)\) as follows \cite{Parry1992}:
\begin{align*}
    (G \wr \mathbb{Z})(z) &= (1 + z^2 (G(z) - 1) (z^2 G(z))^*)^2 G(z) \mathbb{Z}(z G(z)) \\
    &= \left(\frac{1 - z^2}{1 - z^2 G(z)}\right)^2 G(z) \frac{1 + z G(z)}{1 - z G(z)} \\
\end{align*}

Letting \(G = \mathbb{Z}/2\mathbb{Z}\) yields
\[ \mathcal{L}(z) = \left(\frac{1 - z^2}{1 - z^2 - z^3}\right)^2 (1 + z) \frac{1 + z + z^2}{1 - z - z^2} \]

which has a radius of convergence of \(\varphi^{-1}\), where \(\varphi\) is the golden ratio:
\[ \varphi = \frac{1 + \sqrt{5}}{2} = 1.6180 \ldots \]

Hence the growth rate of \(\mathcal{L}\) is given by
\[ \lim_{n \rightarrow \infty} ([z^n] \mathcal{L}(z))^{1/n} = \varphi \]

The singularity at \(z = \varphi^{-1}\) is a simple pole and thus \(\beta = 1\) in our asymptotic expression for the coefficients of a function. Thus our asymptotic coefficient is

\[ \lim_{z \rightarrow r} \frac{\mathcal{L}(z) \left(1 - \frac{z}{r}\right)}{\Gamma(\beta)}^\beta = 3 + \frac{7}{\sqrt{5}} \]

and the lamplighter group grows asymptotically as
\[ [z^n] \mathcal{L}(z) \sim \left( 3 + \frac{7}{\sqrt{5}} \right) \varphi^n \]

We can also introduce a ``lamplighter monoid'' by letting \(g = \varepsilon + t + [t] + [t]t\), which adds the operations of setting a bit to 0 and to 1. This results in \(G(z) = 1 + z + z^3 + z^4\) and
\[ \mathcal{L}_\text{monoid}(z) = \frac{\left(1-z^2\right)^2 \left(z^4+z^3+z+1\right) \left(z \left(z^4+z^3+z+1\right)+1\right)}{\left(1-z \left(z^4+z^3+z+1\right)\right) \left(1-z^2 \left(z^4+z^3+z+1\right)\right)^2} \]

Its radius of convergence is the root of the polynomial
\[ -1 + z + z^2 + z^4 + z^5 \]

which is approximately 0.5518. This singularity is a simple pole and thus \(\beta = 1\). Taking the limit
\[ \lim_{z \rightarrow r} \frac{\mathcal{L}_\text{monoid}(z) \left(1 - \frac{z}{r}\right)}{\Gamma(\beta)}^\beta \]

yields the root of the polynomial
\[ -32 + 1152 z - 14872 z^2 + 30380 z^3 - 30150 z^4 + 5025 z^5 \]

which is approximately 4.8843. Therefore
\[ [z^n] \mathcal{L}_\text{monoid}(z) \sim a r^{-n} \]

where \(a \approx 4.8843\) and \(r \approx 0.5518\). Hence
\[ \lim_{n \rightarrow \infty} ([z^n] \mathcal{L}_\text{monoid}(z))^{1/n} = r^{-1} \approx 1.812 \]

which is slightly higher than the growth rate of \(\mathcal{L}\).

\subsection{Lamplighter programs}

Recall that the grammar which generates our programming language is
\[ E = (\mathcal{G} + [E])^* \]

where \(\mathcal{G} = t + r + l\). Since \((a + b)^* = a^*(ba^*)^*\), we have
\begin{align*}
    E &= (\mathcal{G} + [E])^* \\
    &= \mathcal{G}^* ([E] \mathcal{G}^*)^*
\end{align*}

To avoid double-counting strings of generators in \(\{t, r, l\}^*\) that produce the same lamplighter group element, we would like to represent each group element \(g\) with one of its minimal-length words \(w \in \{t, r, l\}^*\) such that \(\llbracket w \rrbracket = g\). Hence we replace \(\mathcal{G}^*\) with \(\mathcal{L}\):
\[ E = \mathcal{L} ([E] \mathcal{L})^* \]
where \(\mathcal{L}\) contains the minimal-length words, or normal forms, that correspond to each lamplighter group element. Thus the new generating function is given by
\begin{align*}
    E(z) &= \mathcal{L}(z) (z E(z) z \mathcal{L}(z))^* \\
    &= \frac{\mathcal{L}(z)}{1 - z^2 E(z) \mathcal{L}(z)}
\end{align*}

Solving for \(E(z)\) yields
\[ E(z) = \frac{2 \mathcal{L}(z)}{1 + \sqrt{1 - 4 z^2 \mathcal{L}(z)^2}} \]

Its radius of convergence is the smallest root of the polynomial
\[ -1 + 3 z + 7 z^2 - 11 z^4 - 9 z^5 + 2 z^6 + 7 z^7 + 3 z^8 \]

which is approximately \(0.2256\). Hence
\[ \lim_{n \rightarrow \infty} ([z^n] E(z))^{1/n} \approx 4.432 \]

which is smaller than the exponential growth rate of \(5\) we had previously.

Note that \([z^n] E(z) \geq [z^n] \mathcal{L}(z)\), since the growth rate for any programming language capable of representing all group elements is bounded below by the growth rate of the group itself. Thus the exponential growth rate of \(E\) is bounded below by \(\varphi \approx 1.618\). The same applies for \(\mathcal{L}_\text{monoid}\), yielding a slightly higher lower bound of about 1.812.

\subsection{Fixed-shift subsets}

Consider the subset of \(G \wr \mathbb{Z}\) containing all elements with a shift (second component) of \(k \in \mathbb{Z}\):
\[ (G \wr \mathbb{Z})_k = G^{\oplus \mathbb{Z}} \times \{k\} \]

The growth function of this subset is
\[ (G \wr \mathbb{Z})_k(z) = \left(\frac{1-z^2}{1-z^2 G(z)}\right)^2 G(z) (z G(z))^{\lvert k \rvert} \]

Notice that \((G \wr \mathbb{Z})_0 \cong G^{\oplus \mathbb{Z}}\) is a normal subgroup of \(G \wr \mathbb{Z} = G^{\oplus \mathbb{Z}} \rtimes \mathbb{Z}\), as is standard for any semidirect product:
\begin{align*}
    (a_1, b_1) (a_2, 0) (a_1, b_1)^{-1}
    &= (a_1 + \sigma^{b_1} a_2, b_1) (a_1, b_1)^{-1} \\
    &= (a_1 + \sigma^{b_1} a_2, b_1) (-\sigma^{-b_1} a_1, -b_1) \\
    &= (a_1 + \sigma^{b_1} a_2 - \sigma^{b_1} \sigma^{-b_1} a_1, b_1 - b_1) \\
    &= (a_1 + \sigma^{b_1} a_2 - a_1, 0) \\
    &= (\sigma^{b_1} a_2, 0)
\end{align*}

In the lamplighter group, we have
\[ \mathcal{L}_k(z) = \left(\frac{1 - z^2}{1 - z^2 - z^3}\right)^2 (1 + z) (z + z^2)^{\lvert k \rvert} \]

The radius of convergence of this function is \(\rho^{-1}\), where \(\rho\) is the plastic number:
\[ \rho = \frac{\sqrt[3]{108 + 12 \sqrt{69}} + \sqrt[3]{108 - 12 \sqrt{69}}}{6} = 1.3247\ldots \]

Thus \(\rho\) is the exponential growth rate of \(\mathcal{L}_k\). Finally, let \(\mathcal{L}_{k,1}\) and \(\mathcal{L}_{k,0}\) denote the set of \(k\)-shift elements which do and do not, respectively, toggle the final bit. Then
\begin{align*}
    \mathcal{L}_{k,0}(z) &= \left(\frac{1 - z^2}{1 - z^2 - z^3}\right)^2 (z + z^2)^{\lvert k \rvert} \\
    \mathcal{L}_{k,1}(z) &= \mathcal{L}_{k,0}(z) z
\end{align*}

We now start analyzing loop constructs in this programming language to reduce the number of semantically-equivalent programs. In the next section, we study loops that are unreachable and can thus be excluded from the generation process.
\section{Dead loops}

\subsection{After a loop}

In this section, we show that certain loops can never be entered and can thus be eliminated from a program without changing its behavior. For example, the current bit is always 0 immediately after a loop. Hence we can eliminate the second of any two consecutive loops:
\[ [a][b] = [a] \]

More generally, let \(b \in \mathcal{L}_{0,0}\) be a 0-shift element that does not toggle the central bit. Then
\[ b \in \mathcal{L}_{0,0} \Longrightarrow [a]b[c] = [a]b \]

Since \([a]b[c]\) is equivalent to a shorter program (which will be generated earlier in the generation process), we can exclude it. Hence, we exclude all expressions of the form \( [E] \mathcal{L}_{0,0} [E] \) by taking
\begin{align*}
    E &= \mathcal{L} ([E] \mathcal{L})^* \\
    &= \mathcal{L} (\varepsilon + ([E] \mathcal{L})^* [E] \mathcal{L}) \\
    &= \mathcal{L} + \mathcal{L} ([E] \mathcal{L})^* [E] \mathcal{L}
\end{align*}

and remove all \(\mathcal{L}s\) between loops that are in \(\mathcal{L}_{0,0}\):
\[ E = \mathcal{L} + \mathcal{L} ([E] (\mathcal{L} - \mathcal{L}_{0,0}))^* [E] \mathcal{L} \]

Therefore
\begin{align*}
    E(z) &= \mathcal{L}(z) + \mathcal{L}(z) (z^2 E(z) (\mathcal{L}(z) - \mathcal{L}_{0,0}(z)))^* z^2 E(z) \mathcal{L}(z) \\
    &= \mathcal{L}(z) + \frac{z^2 E(z) \mathcal{L}(z)^2}{1 - z^2 E(z) (\mathcal{L}(z) - \mathcal{L}_{0,0}(z))}
\end{align*}

which yields
\[ E(z) = \frac{2 \mathcal{L}(z)}{1 - z^2 \mathcal{L}(z) \mathcal{L}_{0,0}(z) + \sqrt{(1 - 2x \mathcal{L}(z) + z^2 \mathcal{L}(z) \mathcal{L}_{0,0}(z)) (1 + 2x \mathcal{L}(z) + z^2 \mathcal{L}(z) \mathcal{L}_{0,0}(z))}} \]

The radius of convergence of this function is approximately \(0.2409\). Hence
\[ \lim_{n \rightarrow \infty} ([z^n] E(z))^{1/n} \approx 4.150 \]

which is a tighter upper bound than the value of 4.432 we obtained in the previous section.

\subsection{Inside a loop}

Let \(a \in \mathcal{L}_{0,1}\) be a 0-shift lamplighter group element that toggles the central bit. Then
\[ a \in \mathcal{L}_{0,1} \Longrightarrow [a[b]c] = [ac] \]

since the current bit is always 1 at the beginning of a loop body. Thus we exclude all expressions of the form \( [\mathcal{L}_{0,1} [E] E] \) by creating a new nonterminal symbol \(Y\) for loops:
\begin{align*}
    E &= \mathcal{L} + \mathcal{L} (Y (\mathcal{L} - \mathcal{L}_{0,0}))^* Y \mathcal{L} \\
    Y &= [E]
\end{align*}

Expanding the expression inside the loop yields
\begin{align*}
    E &= \mathcal{L} + \mathcal{L} (Y (\mathcal{L} - \mathcal{L}_{0,0}))^* Y \mathcal{L} \\
    Y &= [\mathcal{L} + \mathcal{L} (Y (\mathcal{L} - \mathcal{L}_{0,0}))^* Y \mathcal{L}]
\end{align*}

Replacing the inner \(\mathcal{L}\) with \(\mathcal{L} - \mathcal{L}_{0,1}\) yields
\begin{align*}
    E &= \mathcal{L} + \mathcal{L} (Y (\mathcal{L} - \mathcal{L}_{0,0}))^* Y \mathcal{L} \\
    Y &= [\mathcal{L} + (\mathcal{L} - \mathcal{L}_{0,1}) (Y (\mathcal{L} - \mathcal{L}_{0,0}))^* Y \mathcal{L}]
\end{align*}

Hence
\begin{align*}
    E(z) &= \mathcal{L}(z) + \frac{\mathcal{L}(z)^2 Y(z)}{1 - Y(z) (\mathcal{L}(z) - \mathcal{L}_{0,0}(z))} \\
    Y(z) &= z^2 \left( \mathcal{L}(z) + \frac{\mathcal{L}(z) Y(z) (\mathcal{L}(z) - \mathcal{L}_{0,1}(z))}{1 - Y(z) (\mathcal{L}(z) - \mathcal{L}_{0,0}(z))} \right)
\end{align*}

When solved, the radius of convergence of \(E(z)\) is approximately 0.2443. Hence
\[ \lim_{n \rightarrow \infty} ([z^n] E(z))^{1/n} \approx 4.093 \]

which is even tighter than the last upper bound we obtained. The following graph illustrates the growth sequences we have derived so far. Note that the true growth sequence for semantically distinct programs lies somewhere between that of the lamplighter monoid and that obtained after the dead loop elimination we performed in this section.
\begin{center}
\includegraphics[width=\textwidth]{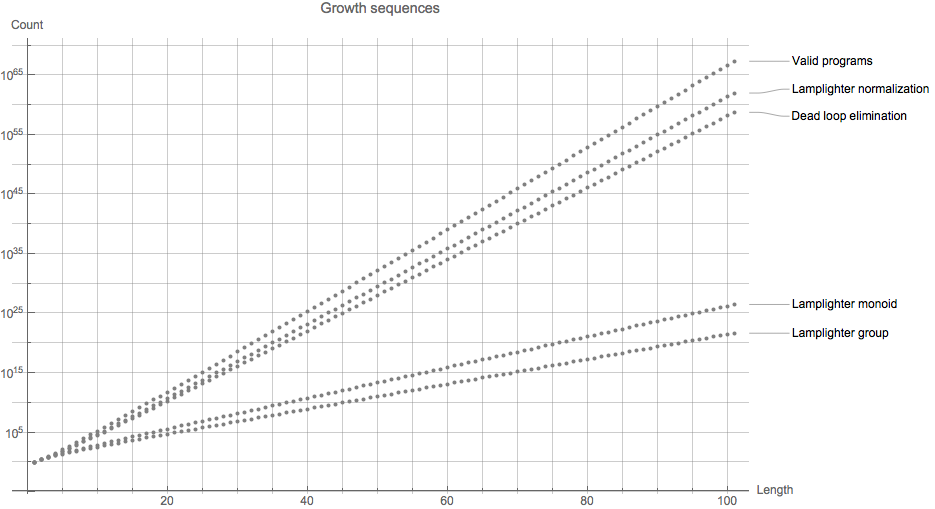}
\end{center}

In the next section, we extend our analysis to loop unrolling, a loop transformation technique that `unrolls' the first iteration of a loop and can be used to perform further analysis.
\section{Loop unrolling}

Consider the simple case of turning a doubly-nested loop into a singly-nested loop:
\[ [[a]] = [a] \]

This transformation preserves semantics because the inner loop is entered whenever the outer loop is entered and the outer loop ends whenever the inner loop ends. Therefore, we can exclude expressions of the form \([[E]] = [Y]\) from the generation process:
\begin{align*}
    E &= \mathcal{L} + \mathcal{L} (Y (\mathcal{L} - \mathcal{L}_{0,0}))^* Y \mathcal{L} \\
    Y &= [\mathcal{L} + (\mathcal{L} - \mathcal{L}_{0,1}) (Y (\mathcal{L} - \mathcal{L}_{0,0}))^* Y \mathcal{L} - Y]
\end{align*}

In general, we are free to unroll the first iteration of a loop when we know that the loop will be executed at least once. By inverting the loop elimination conditions from the previous section, we obtain the following loop unrolling conditions:
\begin{align*}
    b \in \mathcal{L}_{0,1} &\Longrightarrow [a]b[c] = [a]bc[c] \\
    a \in \mathcal{L}_{0,0} &\Longrightarrow [a[b]c] = [ab[b]c]
\end{align*}

This might seem to only make programs longer. However, if we know that the loop can only be executed \textit{at most once}, we can eliminate the rest of the loop and thus inline the loop body:
\begin{align*}
    [a]b[c] = [a]bc[c] = [a]bc \\
    [a[b]c] = [ab[b]c] = [abc]
\end{align*}

A loop that is executed at most once is called a transient loop. This occurs when the current bit is always 0 at the end of the loop body, preventing another iteration. For example:
\begin{align*}
    a \in \mathcal{L}_{0,1} &\Longrightarrow [a] \text{ is transient} \\
    c \in \mathcal{L}_{0,0} &\Longrightarrow [a[b]c] \text{ is transient}
\end{align*}

Hence we have the following reductions:
\begin{align*}
    b \in \mathcal{L}_{0,1}, c \in \mathcal{L}_{0,1} &\Longrightarrow [a]b[c] = [a]bc[c] = [a]bc \\
    b \in \mathcal{L}_{0,1}, e \in \mathcal{L}_{0,0} &\Longrightarrow [a]b[c[d]e] = [a]bc[d]e[c[d]e] = [a]bc[d]e\\
    a \in \mathcal{L}_{0,0}, b \in \mathcal{L}_{0,1} &\Longrightarrow [a[b]c] = [ab[b]c] = [abc] \\
    a \in \mathcal{L}_{0,0}, d \in \mathcal{L}_{0,0} &\Longrightarrow [a[b[c]d]e] = [ab[c]d[b[c]d]e] = [ab[c]de]
\end{align*}

noting that \(\mathcal{L}_{0,a}\mathcal{L}_{0,b} = \mathcal{L}_{0,a + b \bmod 2}\). Since they are provably equivalent to smaller expressions, we can exclude expressions of the following forms from the generation process:

\begin{center}
\begin{tabular}{l l l}
    & After a loop & Inside a loop \\
    Transient type 1 & \([E]\mathcal{L}_{0,1}[\mathcal{L}_{0,1}]\) & \([\mathcal{L}_{0,0}[\mathcal{L}_{0,1}]E]\) \\
    Transient type 2 & \([E]\mathcal{L}_{0,1}[E[E]\mathcal{L}_{0,0}]\) & \([\mathcal{L}_{0,0}[E[E]\mathcal{L}_{0,0}]E]\)
\end{tabular}
\end{center}
\section{Infinite loops} \label{infiniteloops}

Because the halting problem is undecidable, no single algorithm can determine whether a program terminates, for all possible programs. Nonetheless, we can prove that certain classes of programs will (or will not) terminate. This is the goal of termination analysis.

For example, if a loop body consists of a 0-shift lamplighter group element that does not toggle the central bit, the loop will never terminate if entered. This is because the current bit, which is 1 at the beginning of the body, will remain 1 at the end of it. Since the loop condition is always true at the end of the loop body, the loop cannot be escaped.
\[ a \in \mathcal{L}_{0,0} \Longrightarrow [a] \text{ is an infinite loop} \]

Similarly, suppose there is a 0-shift toggling element that follows an inner loop at the end of a loop body. Either the inner loop does not terminate, which \textit{a fortiori} results in an infinite loop, or it terminates when the current bit becomes 0. The current bit is toggled from 0 to 1 as the end of the loop is reached, resulting in an infinite loop:
\[ c \in \mathcal{L}_{0,1} \Longrightarrow [a[b]c] \text{ is an infinite loop} \]

Since they are extensionally equivalent, we can exclude all expressions of the previous two forms from the generation process except for the smallest infinite loop, namely the empty loop \([]\). Suppose we use \(\bot\) denote a point in a program which, if reached, guarantees non-termination. Any instructions before \(\bot\) are absorbed by it since either they do not terminate, which results in \(\bot\), or they do terminate, allowing \(\bot\) to be reached. Similarly, any instructions after \(\bot\) do not affect the non-termination outcome, and are also absorbed by it:
\[ a \bot b = \bot \]

From the previous examples of infinite loops, we have
\begin{align*}
    a \in \mathcal{L}_{0,0} \Longrightarrow& [a] = [\bot] \\
    c \in \mathcal{L}_{0,1} \Longrightarrow& [a[b]c] = [\bot]
\end{align*}

Finally, from the loop unrolling conditions of the previous section we have
\begin{align*}
    a \in \mathcal{L}_{0,0} \Longrightarrow& [a[\bot]b] = [\bot] \\
    b \in \mathcal{L}_{0,1} \Longrightarrow& [a]b[\bot] = \bot
\end{align*}

These rules allow us to propagate the detection of non-termination through a program. Since non-terminating programs are semantically equivalent (namely, they yield non-termination on every input), we can avoid double-counting them in the enumeration process. For example, we have the following sequence of transformations for the program \texttt{[]+[]}:
\begin{align*}
    \texttt{[]+[]} &= \texttt{[]+[\(\bot\)]} \\
    &= \texttt{[]+}\bot \\
    &= \bot
\end{align*}

In the next section we consider a final, broad class of programs which are amenable to analysis of semantic equivalence. This is the class of ``fixed-shift'' programs. It can be shown that a fixed-shift program can only read from and write to a fixed, finite portion of memory. Hence they can be fully analyzed and nontermination is easily detected.

\section{Fixed-shift programs} \label{fixedshift}

Let \(E_i\), where \(i \in \mathbb{Z}\), be defined as follows:
\[ E_i = \mathcal{L}_i + \sum_{j \in \mathbb{Z}} \mathcal{L}_j [E_0] E_{i - j} \]

where \(\mathcal{L}_i\) is the set of strings that represent \(i\)-shift lamplighter group elements. \(E_i\) contains programs that always shift by \(i\) bits (if the program terminates). Each element of \(E_i\) is equivalent to a partial function \((\mathbb{Z}/2\mathbb{Z})^B \rightarrow (\mathbb{Z}/2\mathbb{Z})^B \cup \{\bot\}\) where \(B \in \mathcal{P}_\text{finite}(\mathbb{Z})\) is the finite set of bits that are read or toggled, which can be determined statically, and \(\bot\) denotes non-termination.

A machine configuration consists of its current memory, given by the infinite sequence of bits, and a program counter indicating which instruction is being executed. The set of configurations assumed by a machine executing an expression in \(E_i\) is finite, since \(B\) is finite. Hence it is easy to detect non-termination by detecting whether the machine configuration repeats.

For any set of bits \(B\) and any partial function over those bits \((\mathbb{Z}/2\mathbb{Z})^B \rightarrow (\mathbb{Z}/2\mathbb{Z})^B \cup \{\bot\}\), we can find the shortest program which implements that partial function and exclude all longer, equivalent programs. For example, let \(\mathcal{L}_i^{(k)}\) be the subset of \(\mathcal{L}_i\) that does not toggle bit \(k\). Then we can define \(E_i^{(k)}\) as a subset of \(E_i\) that never toggles bit \(k\):
\[ E_i^{(k)} = \mathcal{L}_i^{(k)} + \sum_{j \in \mathbb{Z}} \mathcal{L}_j^{(k)} [E_0^{(k - j)}] E_{i - j}^{(k - j)} \]

This allows us to extend some of our previous results. For example, we have
\[ b \in E_0^{(0)} \Longrightarrow [a]b[c] = [a]b \]

since \(E_0^{(0)}\) returns to the same bit without ever toggling it. For instance:
\[ n \neq 0 \Longrightarrow [a]r^n[t]l^n[b] = [a]r^n[t]l^n \]

In general, we can replace the role of \(\mathcal{L}_{0,0}\) with \(E_0^{(0)}\), eliminating all such expressions.
\section{Sampling random programs}

One potential application of our analysis is in automated program search. The goal of program search is to search efficiently through the space of possible programs and find ones which satisfy a desired specification. Our analysis reduces the size of the search space by avoiding the generation of duplicate programs with equivalent input-output behavior. This allows techniques like genetic programming \cite{Koza1997}\cite{Koza2000}, which uses an evolutionary algorithm to discover programs that perform well at some predefined task, to find solutions more quickly.

Randomness is an important component of evolutionary algorithms, particularly when generating the initial population. \cite{Flajolet1994} presents a systematic approach to sampling uniformly at random from a set of combinatorial structures. This technique can be applied to any of the combinatorial formulas we derived to specify restricted subsets of our language.

For example, suppose we want to sample a syntactically valid program of size \(n \in \mathbb{N}\) uniformly at random. Recall that the set of syntactically valid programs is generated by the grammar
\[ E = (t + r + l + [E])^* \]

This grammar has the following generating function:
\begin{align*}
    E(z) &= \frac{2}{1 - 3z + \sqrt{(1 - z)(1 - 5z)}} \\
    &= 1 + 3z + 10z^2 + 36z^3 + 137z^4 + 543z^5 + \ldots
\end{align*}

The \(n\)th coefficient of the resulting power series is the number of syntactically valid programs of length \(n\). The coefficients satisfy the following recurrence relation:
\begin{align*}
    c_0 &= 1 \\
    c_1 &= 3 \\
    c_{k+2} &= \frac{3 (2k + 5) c_{k+1} - 5 (k + 1) c_k}{k + 4}
\end{align*}

Before sampling, we generate a table containing \(c_k\) for \(k \in \{0, 1, \ldots, n\}\). To sample a program at random, we use the following algorithm:

\begin{algorithm}[H]
    \caption{Algorithm for sampling a program of length $n$ uniformly at random}
    \begin{algorithmic}
        \Procedure {Sample}{$n$}
            \If {$n = 0$}
                \State \Return $\varepsilon$
            \ElsIf {$\Call{Uniform}{\null} < 3 c_{n-1} / c_n$}
                \State \Return $\Call{Choice}{``\texttt{>}\text{''}, ``\texttt{<}\text{''}, ``\texttt{+}\text{''}} \mathop{\Vert} \Call{Sample}{n-1}$
            \Else
                \State $u \gets  \Call{Uniform}{\null}$
                \State $k \gets 0$
                \State $s \gets c_0 c_{n-2} / (c_n - 3 c_{n-1})$
                \While {$s < u$}
                    \State $k \gets k + 1$
                    \State $s \gets s + c_k c_{n-2-k} / (c_n - 3 c_{n-1})$
                \EndWhile
                \State \Return $``\texttt{[}\text{''} \mathop{\Vert} \Call{Sample}{k} \mathop{\Vert} ``\texttt{]}\text{''} \mathop{\Vert} \Call{Sample}{n-2-k}$
            \EndIf
        \EndProcedure
    \end{algorithmic}
\end{algorithm}

where \(\Vert\) denotes string concatenation, \textsc{Uniform} samples uniformly from the unit interval, and \textsc{Choice} samples uniformly from its arguments. We can adapt this method to sample uniformly from any of the language subsets derived in this paper. Alternatively, one can use the technique of Boltzmann samplers \cite{Duchon2004}, which relax the constraint of generating objects of a strictly fixed size and have a tunable parameter controlling the spread of the distribution.
\section{Distribution of runtimes}

Using the sampling procedure described in the previous section, we can approximate the \textit{runtime distribution} for programs of a given length. That is, we can approximate the probability that a program of length \(\ell\) halts at \(t \in \mathbb{N} \cup \{\infty\}\) steps, where \(\infty\) denotes non-termination.

We count the execution of each of the basic instructions \(\texttt{>}\), \(\texttt{<}\), and \(\texttt{+}\) as 1 step. We also count the expansion of a loop as 1 step:
\begin{align*}
    [a] \rightarrow \begin{cases}
        \varepsilon & \text{current bit is } 0 \\
        a [a] & \text{current bit is } 1
    \end{cases}
\end{align*}

Uniformly sampling \(10^6\) programs for each \(\ell \in \{0,1,\dots,19\}\) and executing them for at most \(t_{\max} = 29\) steps yields the following set of empirical distributions:

\begin{center}
\includegraphics[width=0.8\textwidth]{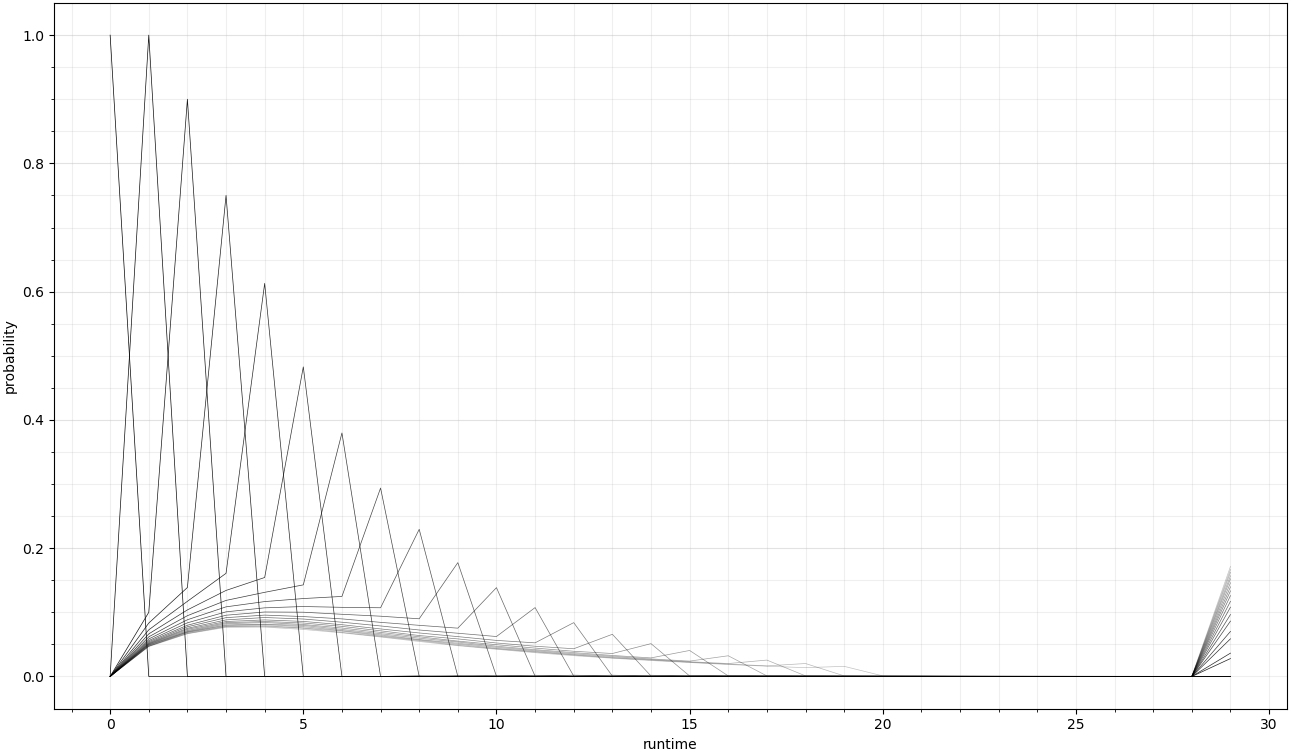}
\end{center}

The empirical distribution is an approximation to the true runtime distribution when censored to lie inside the interval \(\{0, 1, \ldots, t_{\max}\}\). All samples with runtimes greater than \(t_{\max}\), including infinite (non-halting) runtimes, are accumulated at \(t_{\max}\). The true runtime distribution is itself a \textit{mixture} of halting and non-halting components, where the non-halting component is a discrete delta at \(\infty\), while the halting component is a distribution on \(\mathbb{N}\).

A few things can be noted about the shapes of the empirical distributions in the graph above: They initially resemble the function \(f(x) = x \mathrm{e}^{-x}\). However, this function is interrupted by an abrupt increase at \(t = \ell\), followed by an abrupt decrease to negligible probability. This negligible probability persists until \(t_{\max}\) is reached, where the remainder of the halting probability, and the non-halting probability, are accumulated. 

Most programs stop quickly or never halt, in the sense that the set of programs which run for a long time before halting is a set of effectively vanishing probability \cite{Calude2008}. This is consistent with the observed empirical distribution. In particular, programs with \(\ell < t < \infty\) are very rare. This is because \(t > \ell\) means at least one loop in the program must be re-entered (otherwise \(t\) can only be at most \(\ell\)). However, if such a loop \textit{is} re-entered, it is most likely due to a loop invariant that will cause it to be re-entered again ad infinitum, resulting in \(t = \infty\). The shortest programs with \(\ell < t < \infty\) are \texttt{+<+[>]} and \texttt{+>+[<]}, which have \(t = 8\).

The graph below plots the probability accumulated at \(t_{\max}\) against \(\ell\). It increases with \(\ell\) and is nonzero after \(\ell = 3\), which is the length of the shortest nonterminating program, \(\texttt{+[]}\). For \(\ell =\) 100, 200, and 300 we obtained probabilities of 0.2826, 0.2990, and 0.3045 respectively.

\begin{center}
\includegraphics[width=0.8\textwidth]{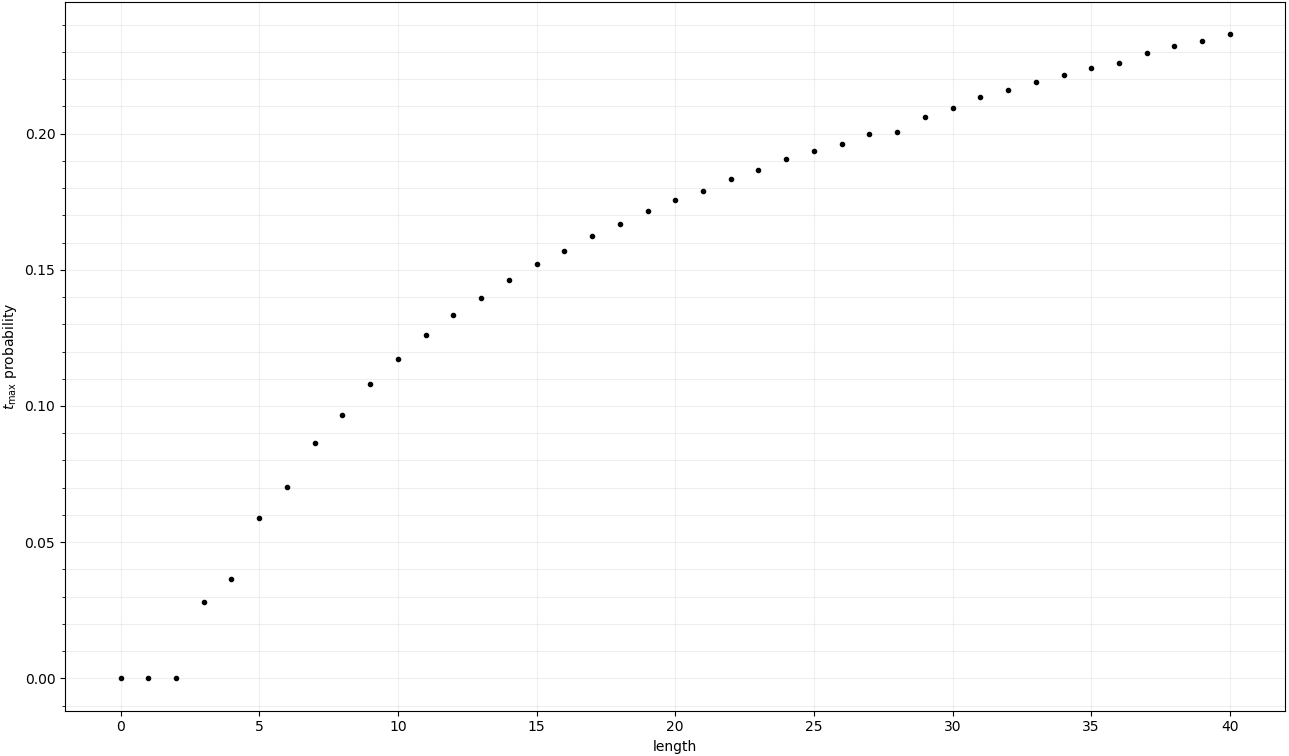}
\end{center}

One can establish a fixed lower bound on the halting probability by finding a set of programs that are guaranteed to halt, and counting how many of those programs have length \(\ell\). Consider for instance the set of programs of the form \(\texttt{[}E\texttt{]}\), where \(E\) is itself any syntactically-valid program. All of these programs are guaranteed to terminate (in 1 step) because the outer loop is not entered since the initial bit is 0. The density of such programs among the set of length \(\ell\) programs is
\begin{align*}
    \frac{c_\ell}{c_{\ell+2}}
    &= \frac{c_\ell}{\frac{3(2\ell+5) c_{\ell+1} - 5(\ell+1) c_\ell}{\ell+4}} \\
    &= \frac{4+\ell}{3(2\ell+5) \frac{c_{\ell+1}}{c_\ell} - 5(\ell+1)}
\end{align*}

Using the asymptotic expression for the number of syntactically-valid programs we derived in section \ref{introduction}, we obtain the following asymptotic density:
\begin{align*}
    \lim_{\ell \rightarrow \infty} \frac{c_\ell}{c_{\ell+2}}
    &= \lim_{\ell \rightarrow \infty} \frac{\frac{5^{\ell+3/2}}{2\sqrt{\pi \ell^3}}}{\frac{5^{\ell+2+3/2}}{2\sqrt{\pi (\ell+2)^3}}} \\
    &= \lim_{\ell \rightarrow \infty} \frac{(\ell+2)^{3/2}}{25 \ell^{3/2}} \\
    &= \frac{1}{25} \\
    &= 0.04
\end{align*}

Similarly, one can establish a fixed upper bound by finding a set of programs that are guaranteed to never halt, and counting how many of those programs have length \(\ell\). Section \ref{infiniteloops} describes precisely such a set of programs that are guaranteed to never terminate. For example, consider the set of programs of the form \(\texttt{+[]}E\). They all change the initial bit to 1 and then loop forever. Using the same calculation as before, we obtain the following asymptotic density:
\begin{align*}
    \lim_{\ell \rightarrow \infty} \frac{c_\ell}{c_{\ell+3}}
    &= \frac{1}{5^3}
\end{align*}

To make the previous bounds unconditional on the state of the initial bit, we can replace \(\texttt{[}E\texttt{]}\) with \(\texttt{[+][}E\texttt{]}\) and \(\texttt{+[]}E\) with \(\texttt{[]+[]}E\), leading to an asymptotic density of \(5^{-5}\) in both cases. It is known that almost all programs in the lambda calculus are strongly normalizing, and almost all programs in the SKI combinator calculus are \textit{not} strongly normalizing \cite{David2009}. Based on the previous results, neither is the case with lamplighter programs.

\subsection{Preliminary combinatorial analysis}

We now pursue a tighter lower bound on the probability that a program of length \(\ell\) has a runtime of \(t\) steps. Recall that the generating function \(f : \mathbb{C} \rightarrow \mathbb{C}\) of a sequence \(a : \mathbb{N} \rightarrow \mathbb{C}\) is
\[f(z) = \sum_{n \in \mathbb{N}} a_n z^n\]

Let \(L\) be the set of words generated by the standard generators of the lamplighter group, as described in section \ref{lamplighter}. Since there are 3 such generators, we have
\begin{align*}
    L(z) &= (3z)^* \\
    &= \frac{1}{1-3z} \\
    [z^n] L(z) &= 3^n
\end{align*}

Let \(L_k\) be the set of lamplighter group words that shift the lamplighter by \(k\). Then
\[[z^n] L_k(z) = \binom{n}{k}_2\]

where the RHS is the entry in row \(n\) and column \(k\) of the trinomial triangle. For instance, \(L_0\) is the set of words that return the lamplighter to its initial position. The number of length-\(n\) words in \(L_0\) is the \(n\)th central trinomial coefficient (sequence A002426 on OEIS):
\begin{align*}
    [z^n] L_0(z) = \sum_{i=0}^n \binom{n}{i} \binom{i}{n - i}
    &\sim \frac{3^{n+1/2}}{2 \sqrt{\pi n}}
\end{align*}

Thus \(L_0\) has the generating function
\begin{align*}
    L_0(z) &= \frac{1}{\sqrt{(1+z)(1-3z)}} \\
    &= 1 + z + 3z^2 + 7z^3 + 19z^4 + 51z^5 + 141z^6 + 393z^7 + \dots
\end{align*}

Let \(L_k^-\) be the subset of \(L_k\) such that the lamp at \(k\)---or, equivalently in number, the lamp at the final position of the lamplighter---is not toggled. Let \(L_k^+\) be the counterpart of \(L_k^-\) that does toggle this lamp. Then \(L_k = L_k^- + L_k^+\). Furthermore, \(L_k^+ = L_k^- T N\), where \(T\) is a toggling instruction and \(N\) is the set of words that return the lamplighter to the initial position, but never toggle the lamp at this position over the course of its action. We know that
\[L_0 = N (T N)^*\]

That is, a zero-shift word is composed of zero-shift words that do not toggle the central lamp, interposed with toggling instructions. Thus
\begin{align*}
    L_0(z)
    &= \frac{N(z)}{1 - z N(z)} \\
    N(z)
    &= \frac{L_0(z)}{1 + z L_0(z)} \\
    &= \frac{1}{z+\sqrt{(1+z)(1-3z)}} \\
    &= 1 + 2z^2 + 3z^3 + 8z^4 + 16z^5 + 46z^6 + 114z^7 + \dots
\end{align*}

Therefore
\begin{align*}
    L_k(z)
    &= L_k^-(z) + L_k^+(z) \\
    &= L_k^-(z) + L_k^-(z) z N(z) \\
    &= L_k^-(z) (1 + z N(z)) \\
    L_k^-(z) &= \frac{L_k(z)}{1 + z N(z)} \\
    &= L_k(z) \frac{1 + z L_0(z)}{1 + 2 z L_0(z)}
\end{align*}

In particular, we have
\begin{align*}
    L_0^-(z) &= L_0(z) \frac{1 + zL_0(z)}{1 + 2zL_0(z)} \\
    &= \frac{z + \sqrt{(1+z)(1-3z)}}{\sqrt{(1+z)(1-3z)} (2z + \sqrt{(1+z)(1-3z)})} \\
    L^-(z) &= \sum_{k \in \mathbb{Z}} L_k(z) \frac{1 + zL_0(z)}{1 + 2zL_0(z)} \\
    &= L(z) \frac{1 + zL_0(z)}{1 + 2zL_0(z)} \\
    &= \frac{z + \sqrt{(1+z)(1-3z)}}{(1-3z) (2z + \sqrt{(1+z)(1-3z)})}
\end{align*}

where \(L^-\) is the set of all lamplighter words that do not toggle the lamp at the final position of the lamplighter (which is equinumerous with the set of words that do not toggle the lamp at the \textit{initial} position of the lamplighter).

\subsection{Bounds for the runtime distribution}

Consider the set of programs of the following form:
\[\Omega = (\varepsilon + (L^- - L_0^-)[E]) (L_0^-[E])^* L\]

Notice that none of the outer loops in a program of this form are ever entered, so execution proceeds linearly from left to right. First, the lamplighter either (1) does nothing or (2) performs a sequence of lamp manipulations that leave the final lamp untoggled, thus avoiding the first \([E]\). We exclude manipulations in \(L_0^-\) to avoid double-counting with the starred term. Second, the lamplighter performs manipulations that return it to the same position and leave the lamp at this position untoggled, thus avoiding the next outer loop. This second step is repeated an arbitrary number of times. Finally, the lamplighter performs an arbitrary sequence of manipulations.

The syntactic structure and execution semantics of programs in \(\Omega\) are simple enough to make it amenable to length and runtime analysis. We derive a bivariate generating function for \(\Omega\) that has both the length of a program \(z\) and its runtime \(w\) as variables:
\begin{align*}
    \Omega(z, w)
    &= (1 + (L^-(z w) - L_0^-(z w)) w z E(z) z) (L_0^-(z w) w z E(z) z)^* L(z w) \\
    &= \frac{L(z w) (1 + w z^2 (L^-(z w) - L_0^-(z w)) E(z))}{1 - w z^2 L_0^-(z w) E(z)}
\end{align*}

Notice that we use \(z w\) as the input to the generating functions of all \(L\) terms. This is because the length and runtime of lamplighter words are equal. Furthermore, we do not include \(w\) as an input to the generating functions of any \(E\) terms because they are inside the outer loops, and therefore never executed. We only care about their syntactic length.

A lower bound on the number of length-\(\ell\) programs that terminate at time \(t\) is \([z^\ell] [w^t] \Omega(z, w)\). Thus a lower bound on the fraction of length-\(n\) programs which terminate at time \(t\) is just this quantity divided by \([z^\ell] E(z)\). Below is a plot of the lower bound given by \(\Omega\) over different values of \(\ell\) and \(t\):

\begin{center}
\includegraphics[width=0.8\textwidth]{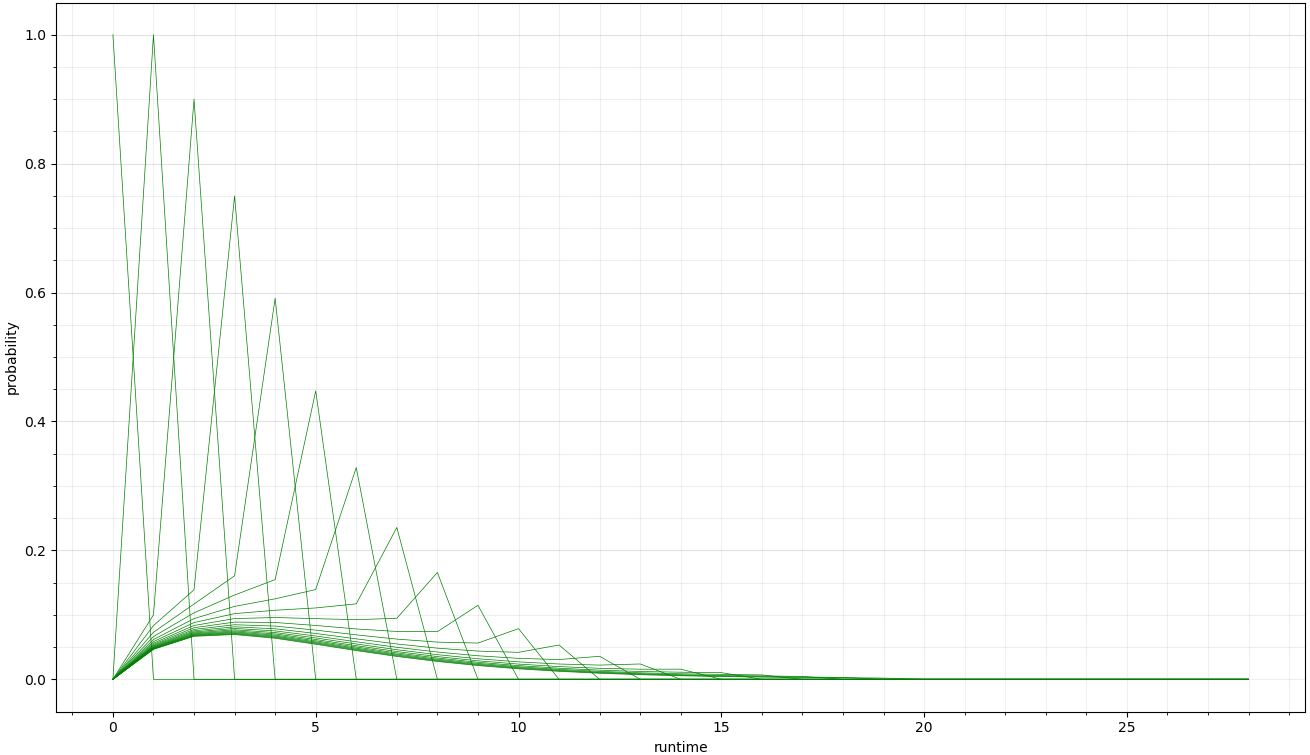}
\end{center}

These distributions closely match the empirical distributions obtained through random sampling and execution of programs. The uptick in probability that occurs when \(t = \ell\) is due to the \(3^\ell\) lamplighter words of length \(\ell\). This can be seen by replacing \(\Omega\) with \(\Omega - L\):

\begin{center}
\includegraphics[width=0.8\textwidth]{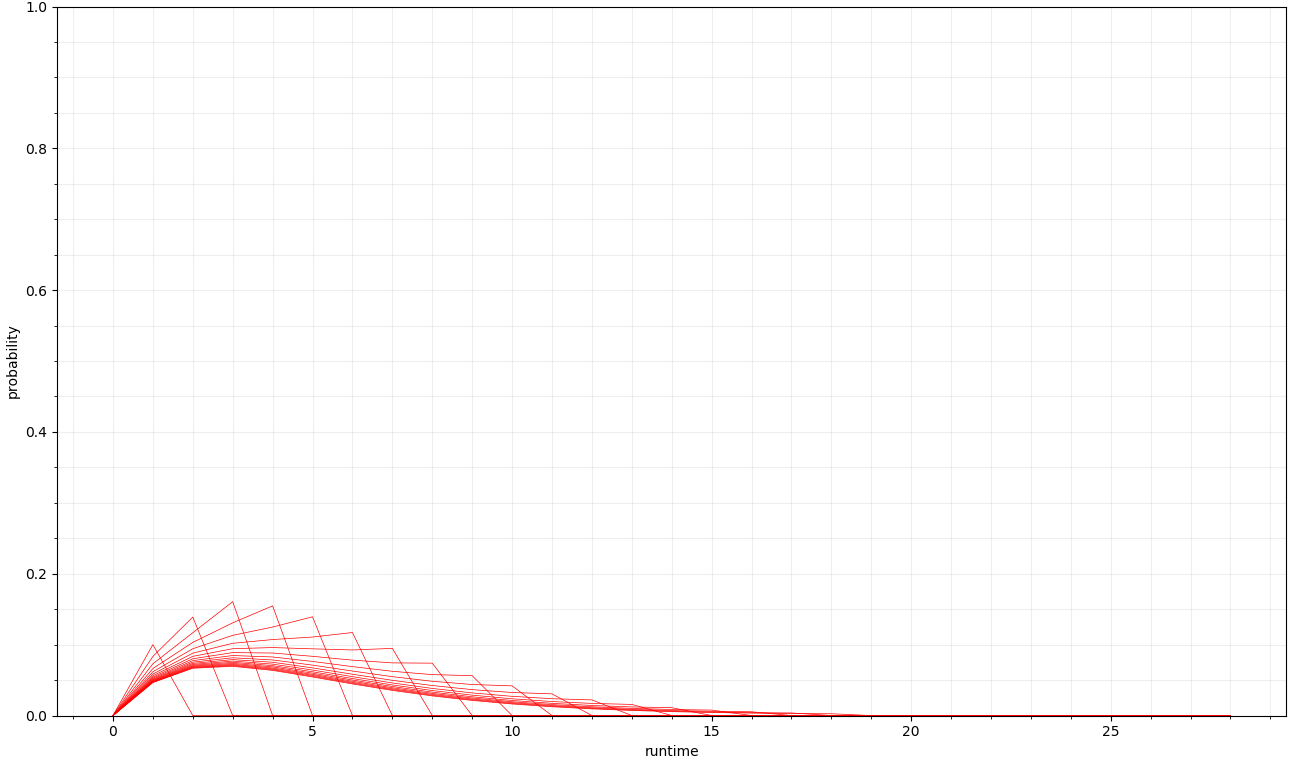}
\end{center}

Every program in \(\Omega - L\) contains at least one loop, so it is always the case that \(t\) is strictly less than \(\ell\) (since no loops are ever entered). Furthermore,
\begin{align*}
    f(z) = \sum_{n \in \mathbb{N}} a_n z^n \Longrightarrow f(1) = \sum_{n \in \mathbb{N}} a_n
\end{align*}

Hence a lower bound on the number of terminating length-\(\ell\) programs is \([z^\ell] \Omega(z, 1)\), where
\begin{align*}
\Omega(z, 1) = 1 + 3z + 10z^2 + 35z^3 + 129z^4 + 490z^5 + 1912z^6 + \dots
\end{align*}

Thus \([z^\ell] \Omega(z, 1) / [z^\ell] E(z)\) lower-bounds the halting probability of length-\(\ell\) programs. As \(\ell \rightarrow \infty\), this bound approaches about 0.35. More precisely, recall from section \ref{introduction} that if
\[ f(z) = \frac{A(z) + B(z) (1 - z/r)^{-\beta}}{z^\alpha} \]

where \(r\) is the radius of convergence of \(f\), the radius of convergence of \(A\) and \(B\) is greater than \(r\), and \(B(r) \neq 0\), then
\[ [z^n] f(z) \sim \frac{B(r)}{r^\alpha \Gamma(\beta)} n^{\beta-1} r^{-n} \]

Let \(f(z) = \Omega(z, 1)\). Due to a \(\sqrt{1-5z}\) term, the nearest singularity to the origin is \(z = 1/5\), so \(r = 1/5\). This singularity is created by a square root so \(\beta = -1/2\). Since \(\beta < 0\),
\[ f(r) = \frac{A(r)}{r^\alpha} \]

Hence
\begin{align*}
    \lim_{z \rightarrow r} (f(z) - f(r)) (1 - z/r)^\beta
    &= \lim_{z \rightarrow r} ( f(z) (1-z/r)^\beta - f(r) (1-z/r)^\beta ) \\
    &= \lim_{z \rightarrow r} \left( \frac{A(z) (1-z/r)^\beta + B(z)}{z^\alpha} - \frac{A(r)}{r^\alpha} (1-z/r)^\beta \right) \\
    &= \lim_{z \rightarrow r} \left( \left( \frac{A(z)}{z^\alpha} - \frac{A(r)}{r^\alpha} \right) (1-z/r)^\beta + \frac{B(z)}{z^\alpha} \right) \\
    &= \lim_{z \rightarrow r} \left( \frac{A(z)}{z^\alpha} - \frac{A(r)}{r^\alpha} \right) (1-z/r)^\beta + \lim_{z \rightarrow r} \frac{B(z)}{z^\alpha} \\
    &= \frac{B(r)}{r^\alpha}
\end{align*}

Therefore
\begin{align*}
    [z^n] f(z) &\sim \frac{B(r)}{r^\alpha \Gamma(\beta)} n^{\beta-1} r^{-n} \\
    &= \frac{\lim_{z \rightarrow r} (f(z) - f(r)) (1 - z/r)^\beta}{\Gamma(\beta)} n^{\beta - 1} r^{-n} \\
    &= \frac{\lim_{z \rightarrow r} \left(f(z) - \frac{845 - 5 \sqrt{3}}{218}\right) (1 - 5z)^{-1/2}}{\Gamma(-1/2)} n^{-1/2 - 1} 5^n \\
    &= \frac{-\frac{630 \sqrt{5}+270 \sqrt{15}}{\left(6+11 \sqrt{3}\right)^2}}{-2 \sqrt{\pi}} n^{-3/2} 5^n \\
    &= \frac{315 \sqrt{5}+135 \sqrt{15}}{\left(6+11\sqrt{3}\right)^2 \sqrt{\pi}} n^{-3/2} 5^n \\
    &= \frac{8025 \sqrt{5} + 1365 \sqrt{15}}{11881 \sqrt{\pi}} n^{-3/2} 5^n
\end{align*}

Thus
\begin{align*}
    \frac{[z^\ell] \Omega(z, 1)}{[z^\ell] E(z)}
    &\sim \frac{\frac{8025 \sqrt{5} + 1365 \sqrt{15}}{11881 \sqrt{\pi}} n^{-3/2} 5^n}{\frac{5^{3/2}}{2 \sqrt{\pi}} n^{-3/2} 5^n} \\
    &= \frac{546 \sqrt{3}+3210}{11881} \\
    &\approx 0.35
\end{align*}

In fact, using the asymptotic expression \(g(w)\) for \([z^\ell] \Omega(z, w) / [z^\ell] E(z)\) and plotting its coefficients \([w^t] g(w)\) yields the following runtime distribution for the limit where \(\ell \rightarrow \infty\):

\begin{center}
\includegraphics[width=0.8\textwidth]{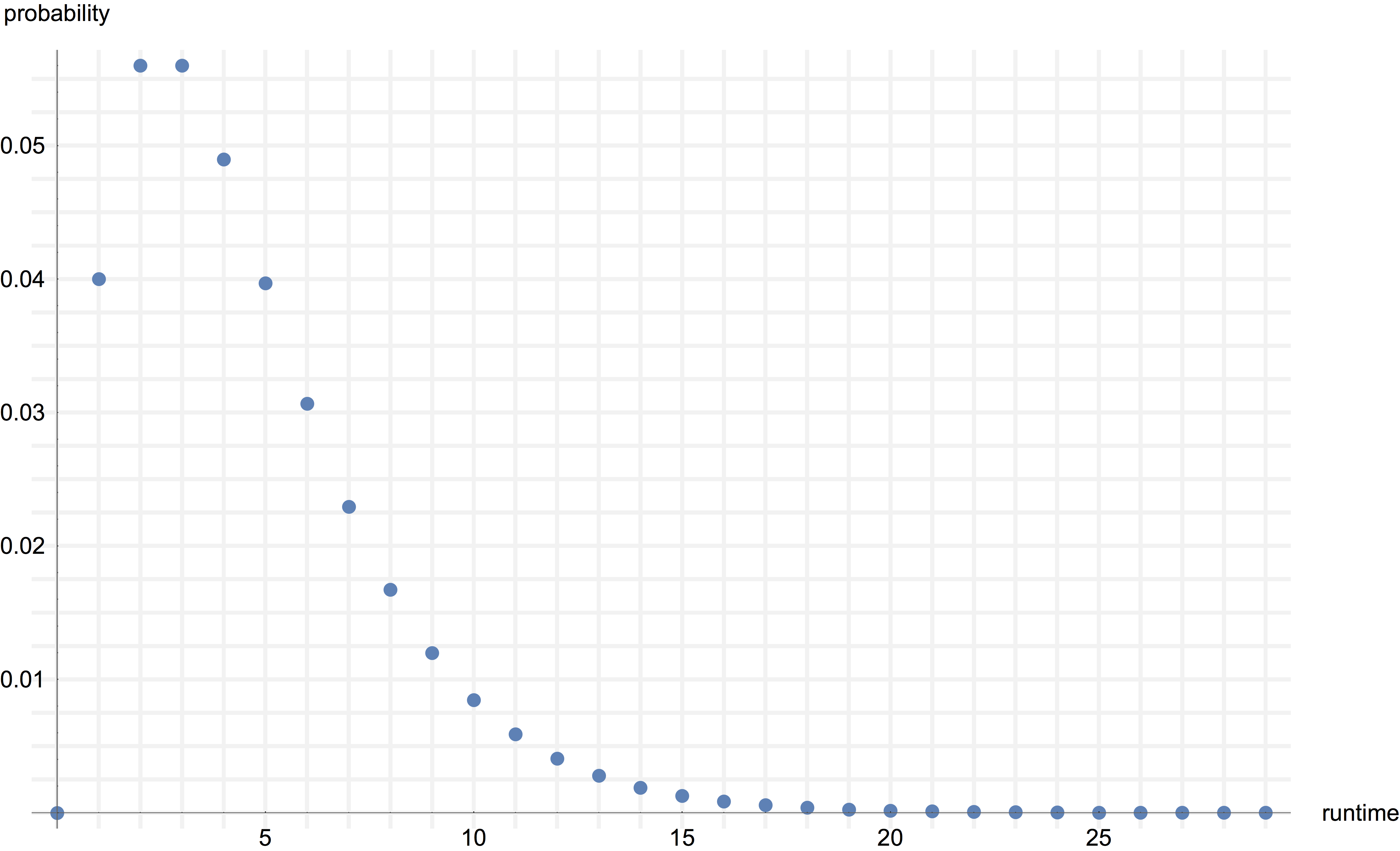}
\end{center}

Similarly, the expression
\begin{align*}
    \Lambda &= (L^+ + E[E] L_0^+) [L_0^- + E[E]L_0^+] E \\
    \Lambda(z) &= z^3 + 5 z^4 + 29 z^5 + 134 z^6 + 630 z^7 + 2823 z^8 + 12689 z^9 + \dots
\end{align*}

yields a lower bound on the \textit{non}-halting probability. This is because the current bit is always 1 when the \([L_0^- + E[E]L_0^+]\) is encountered. As we saw in section \ref{infiniteloops}, \([L_0^-] = [E[E]L_0^+] = [\bot]\). Thus \(1 - [z^\ell] \Lambda(z) / [z^\ell] E(z)\) yields an upper bound on the halting probability. Like \(\Omega(z, 1)\), \(\Lambda(z)\) has a square-root singularity at \(z = 1/5\). Repeating the same procedure we did for \(\Omega\) yields
\begin{align*}
    [z^n] \Lambda(z)
    &\sim \frac{15 \sqrt{15} - 5 \sqrt{5}}{96 \sqrt{\pi}} n^{-3/2} 5^n
    \\
    \frac{[z^\ell] \Lambda(z)}{[z^\ell] E(z)}
    &\sim \frac{3 \sqrt{3} - 1}{48} \\
    &\approx 0.09
\end{align*}

The upper and lower bounds we just derived for the halting probability of a length-\(\ell\) program are shown in the graph below:

\begin{center}
\includegraphics[width=0.8\textwidth]{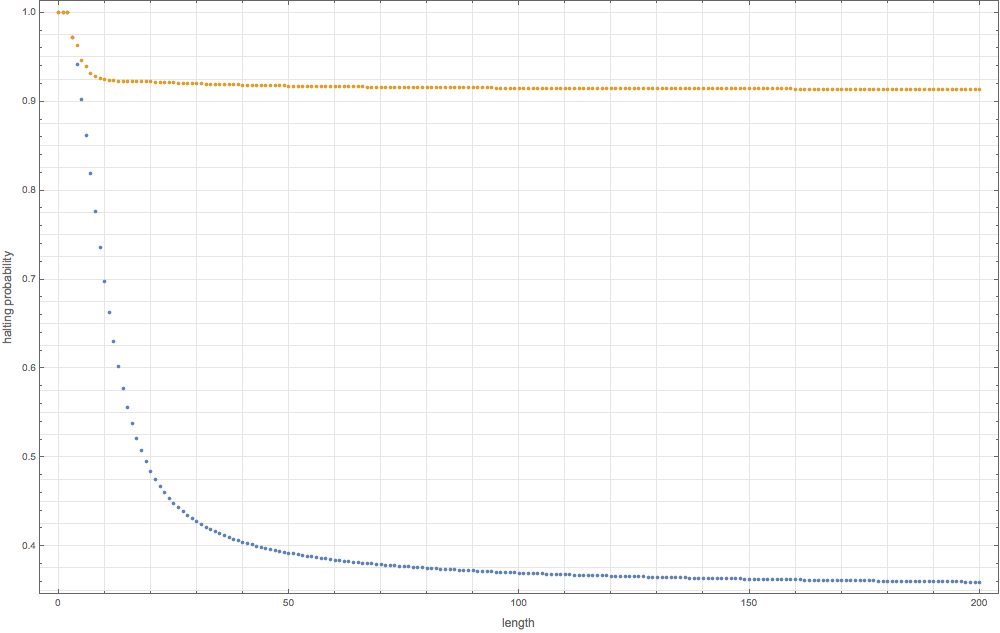}
\end{center}

\cite{Hamkins2005} proves that the halting problem is decidable on a set of programs of asymptotic probability one, given a particular model of Turing machines. Similarly, one can search for sets of lamplighter programs that are decidable, perhaps in polynomial time, and have high asymptotic probability, perhaps equal to one. An example of a large, decidable set of lamplighter programs is the set of fixed-shift programs \(E_i\) described in section \ref{fixedshift}.

A way to characterize the generating function of \(E_i\) is by using \(L_i\) directly:
\begin{align*}
    E_0(z) &= (z E_0(z) z)^* L_0(z (z E_0(z) z)^*) \\
    &= -\frac{1}{\sqrt{1-2z-3z^2-2z^2E_0(z)+2z^3E_0(z)+z^4E_0(z)^2}} \\
    E_i(z) &= (z E_0(z) z)^* L_i(z (z E_0(z) z)^*)
\end{align*}

where \((z E_0(z) z)^*\) represent sequences of terms of the form \([E_0]\), which can be placed anywhere in a lamplighter word.

\subsection{Approximate execution}

We found that the runtime distribution is well-approximated for large \(\ell\) by a modified execution procedure. If a loop is entered, we execute its body either \textit{once} or an \textit{infinite} number of times, with equal probability. That is, we replace \([a]\) with either \(a\) or \(\bot\). In doing so, we are essentially performing the approximation \([a] \approx \varepsilon + a + \bot\). This approximation can be justified by the fact that, just as a program running for a long but finite time is unlikely, executing a loop for a large but finite number of iterations is unlikely. Increasing the maximum number of iterations before a loop is heuristically judged to be non-terminating, i.e. \([a] \approx \varepsilon + a + a^2 + a^3 + \dots + \bot\), increases the accuracy of this approximation.

Notice the runtime for approximate execution is upper-bounded by the length of the program, since every loop is expanded at most once (or infinitely, in which case execution is immediately stopped and the program is judged to be non-terminating). The plot below shows the empirical runtime distributions for \(\ell = 300\) under exact and approximate execution:

\begin{center}
\includegraphics[width=0.8\textwidth]{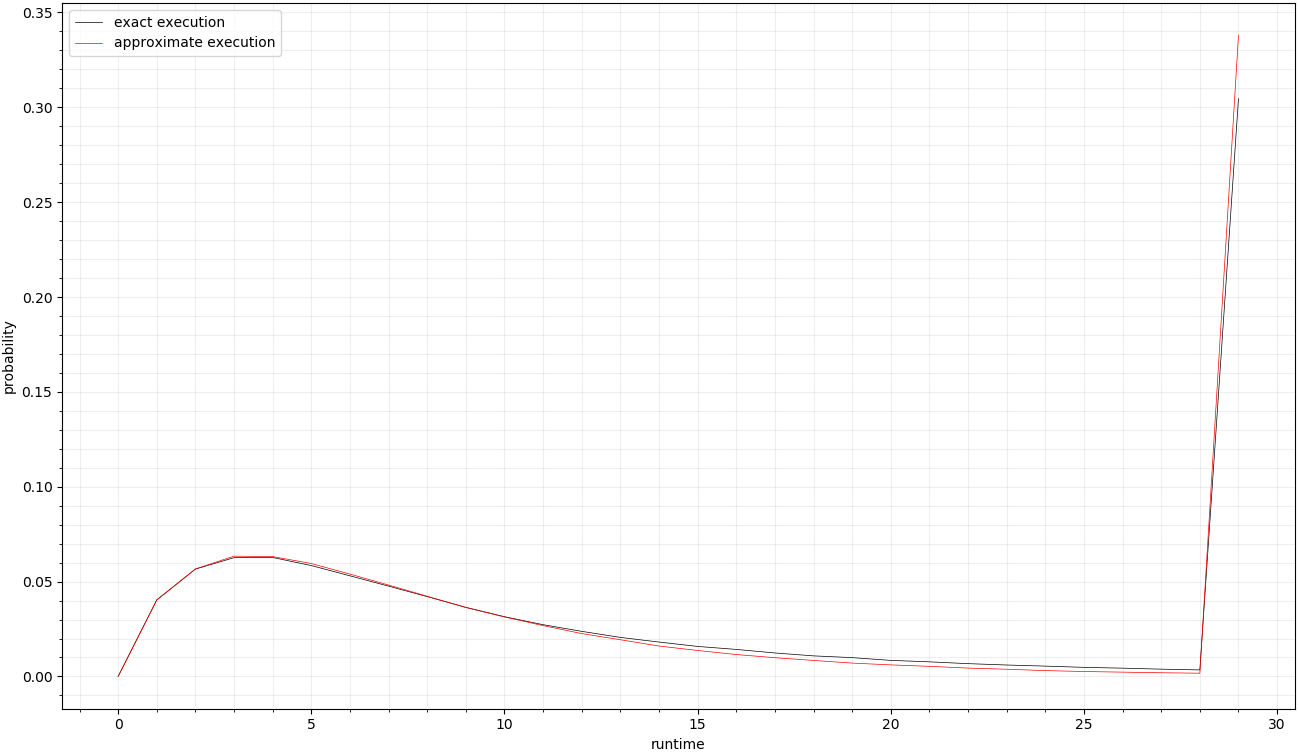}
\end{center}

They are almost identical, suggesting the latter is a good way to approximate the true runtime distribution, even when the program length is very large. There could be other execution models that also approximate exact execution with better space and time complexity.

The runtime distribution may itself change when programs are sampled uniformly from subsets of the original language, like those described throughout this paper. In particular, we expect that the proportion of trivial runtimes decreases under these distributions. Due to space constraints, we leave these problems to future work.

\section{Conclusion}

We have analyzed the semantics of a programming language that manipulates an infinite sequence of bits using function composition and iteration. Our goal was to find classes of programs which are semantically equivalent, that is, yield the same outputs on every possible input.

To do this, we studied the relationship between the language's instructions for manipulating memory and the lamplighter group. We performed reachability analysis to statically determine when loops are entered and how they can be transformed while preserving semantics. 

By deriving combinatorial generating functions for restricted subsets of this language and analyzing their asymptotic behavior, we establish increasingly tight lower and upper bounds on the growth rate of semantically distinct programs. In the previous section, we also explored the distribution of runtimes when a program of fixed length is sampled from the language.

We hope that this combinatorial analysis will be extended to more complex classes of equivalent programs, yielding further reductions. In particular, we would like to find tighter upper and lower bounds on the growth rate of the set of semantically distinct programs.

\bibliographystyle{unsrt}
\bibliography{references} 

\section*{Python code for program sampling and execution}


\begin{lstlisting}[language=python,  basicstyle=\footnotesize\ttfamily,numbers=left]
import numpy as np
import matplotlib.pyplot as plt
from collections import deque

def get_string(program):
  if len(program) == 0:
    return ''
  elif isinstance(program[0], list):
    return '[' + get_string(program[0]) + ']' + get_string(program[1:])
  else:
    return program[0] + get_string(program[1:])

def get_coeffs(length):
  coeffs = [None] * (length + 1)
  coeffs[0] = 1
  coeffs[1] = 3
  for k in range(length - 1):
    coeffs[k + 2] = (3*(2*k + 5) * coeffs[k + 1] - 5*(k+1) * coeffs[k]) // (k + 4)
  return np.array(coeffs)

def sample_program(length, coeffs):
  if length == 0:
    return []
  elif np.random.uniform() < 3 * coeffs[length - 1] / coeffs[length]:
    return [np.random.choice(['>', '<', '+'])] + sample_program(length - 1, coeffs)
  else:
    u = np.random.uniform()
    k = 0
    s = coeffs[0] * coeffs[length - 2] / (coeffs[length] - 3 * coeffs[length - 1])
    while s < u:
      k += 1
      s += coeffs[k] * coeffs[length - 2 - k] / (coeffs[length] - 3 * coeffs[length - 1])
    return [sample_program(k, coeffs)] + sample_program(length - 2 - k, coeffs)

def execute(program, max_time):
  head = 0
  tape = set()
  program = deque(program)
  for time in range(max_time):
    if len(program) == 0:
      break
    instruction = program.popleft()
    if instruction == '>':
      head += 1
    elif instruction == '<':
      head -= 1
    elif instruction == '+':
      tape ^= {head}
    elif isinstance(instruction, list) and head in tape:
      program.appendleft(instruction)
      program.extendleft(instruction)
  return time, head, tape

def execute_approx(program, max_time):
  head = 0
  tape = set()
  program = deque(program)
  for time in range(max_time):
    if len(program) == 0:
      break
    instruction = program.popleft()
    if instruction == '>':
      head += 1
    elif instruction == '<':
      head -= 1
    elif instruction == '+':
      tape ^= {head}
    elif isinstance(instruction, list) and head in tape:
      if np.random.uniform() < .5:
        program.extendleft(instruction)
      else:
        return max_time - 1
  return time, head, tape

def enum_programs(length):
  if length == 0:
    yield []
  else:
    for remainder in enum_programs(length - 1):
      yield ['>'] + remainder
      yield ['<'] + remainder
      yield ['+'] + remainder
    for body_length in range(length - 1):
      for body in enum_programs(body_length):
        for remainder in enum_programs(length - 2 - body_length):
          yield [body] + remainder

max_length = 20
max_time = 30
trials = 10**5

coeffs = get_coeffs(max_length)
counts = np.zeros((max_length, max_time))
counts_approx = np.zeros((max_length, max_time))

for length in range(max_length):
  print('Counting runtimes for programs of length {}'.format(length))
  counts[length] = np.bincount([
      execute(sample_program(length, coeffs), max_time)[0] 
      for trial in range(trials)
  ], minlength=max_time)
  counts_approx[length] = np.bincount([
      execute_approx(sample_program(length, coeffs), max_time)[0] 
      for trial in range(trials)
  ], minlength=max_time)

plt.xlabel('runtime')
plt.ylabel('probability')
plt.plot(counts.T / counts.sum(1), color='black', linewidth=.5)
plt.plot(counts_approx.T / counts_approx.sum(1), color='red', linewidth=.5)
plt.plot(3**np.arange(max_length) / coeffs, color='red', linewidth=.5)
plt.tight_layout()
plt.minorticks_on()
plt.grid(True, which='both', alpha=.2)
plt.show()
\end{lstlisting}

\end{document}